\documentclass[aip,twocolumn,a4paper,showkeys,showpacs,apl,preprint,floatfix]{revtex4-1}
\usepackage{amsmath}
\usepackage{amsfonts}
\usepackage{amssymb}
\usepackage{graphicx}

\newcommand{\NTT}{\affiliation{NTT Basic Research Laboratories, NTT Corporation, 3-1 Morinosato-Wakamiya, Atsugi, Kanagawa, 243-0198, Japan}}
\newcommand{\NII}{\affiliation{National Institute of Informatics, 2-1-2 Hitotsubashi, Chiyoda-ku, Tokyo 101-8430, Japan}}
\newcommand{\Bin}{\mathbf{B}_{\parallel}}

\begin{document}
\author{Hiraku~ Toida}\email{toida.hiraku@lab.ntt.co.jp}\NTT
\author{Yuichiro~ Matsuzaki}\NTT
\author{Kosuke~ Kakuyanagi}\NTT
\author{Xiaobo~ Zhu}\email[Current address: ]{The Institute of Physics, Chinese Academy of Sciences P.O.Box 603, Beijing 100190, China}\NTT 
\author{William~J.~ Munro}\NTT
\author{Kae~ Nemoto}\NII
\author{Hiroshi~ Yamaguchi}\NTT
\author{Shiro~ Saito}\NTT

\title{Electron paramagnetic resonance spectroscopy using a dc-SQUID magnetometer directly coupled to an electron spin ensemble}

\begin{abstract}
We demonstrate electron spin polarization detection and electron paramagnetic resonance (EPR) spectroscopy using a direct current superconducting quantum interference device (dc-SQUID) magnetometer.
Our target electron spin ensemble is directly glued on the dc-SQUID magnetometer that detects electron spin polarization induced by a external magnetic field or EPR in micrometer-sized area.
The minimum distinguishable number of polarized spins and sensing volume of the electron spin polarization detection and the EPR spectroscopy are estimated to be $\sim$$10^6$ and $\sim$$10^{-10}$ $\mathrm{cm}^{3}$ ($\sim$0.1 pl), respectively.
\end{abstract}

\maketitle

Electron paramagnetic resonance EPR spectroscopy is a widely-used method to obtain material properties such as the Land\'{e} factor of electron spins in various materials \cite{Schweiger2001}.
Conventional EPR spectrometers use a microwave cavity as a detector of permeability change induced by electron spin polarization \cite{Schweiger2001}.
Recent technological progress in superconducting circuits including Josephson junctions enables us to use these superconducting devices as a sensitive detector of permeability at low temperatures.
Using superconducting coplanar waveguide resonators, EPR spectroscopy of various materials, such as nitrogen vacancy (NV) centers \cite{Kubo2010} and nitrogen substitution (P1) centers \cite{Schuster2010} in diamond, chromium doped aluminum oxide \cite{Schuster2010}, and erbium impurities in yttrium orthosilicate (Y$_2$SiO$_5$, YSO) \cite{Bushev2011} has been demonstrated. 
By hybridizing a superconducting resonator and a superconducting transmon qubit, highly sensitive EPR spectroscopy was also demonstrated \cite{Kubo2012}.
In these devices, coplanar waveguide resonators play the role of detectors of spin polarization \cite{Kubo2010,Schuster2010,Bushev2011} as the microwave cavity does in the conventional EPR spectrometers.
Recently, metallic coplanar waveguides are also used \cite{Wiemann2015} to perform on-chip ESR spectroscopy in wider temperature and magnetic field range.
On the other hand, EPR spectroscopy has also been performed with a pickup coil and a direct current superconducting quantum interference device (dc-SQUID) amplifier \cite{Chamberlin1979, Sakurai2013}.
In this method, the change of magnetization induced by EPR is detected by a pickup coil and the information is transferred to the dc-SQUID amplifier placed far from the sample to avoid high magnetic fields.
In these EPR spectrometers using superconducting devices, spatial resolution is limited by the detector size, which is typically millimeter scale.
To improve the spatial resolution, a field gradient is used in the case of conventional EPR spectrometers.
In such a case, the spatial resolution is limited to about 0.1 mm due to linewidth of the spectrum and strength of the field gradient.

In this Letter, we demonstrate a novel method for EPR spectroscopy using a dc-SQUID magnetometer.
We use a direct magnetic coupling between a dc-SQUID and electron spins to detect the electron spin polarization induced by EPR.
This scheme has promising characteristics as follows:
Dc-SQUIDs are known to be a sensitive magnetometer with the sensitivity of $\sim$$10^{-15}$ T/$\sqrt{\mathrm{Hz}}$ \cite{Drung2007}, which could provide us with more reliable detection ability of the spins.
The spatial resolution and the sensitivity depend on the dc-SQUID loop size, because the dc-SQUID detects the magnetic field penetrating the dc-SQUID loop.
The dc-SQUID size can be as small as tens of nanometers \cite{Vasyukov2013}, thus nanoscale EPR spectroscopy is in principle possible.
Furthermore, a superconducting flux qubit \cite{Mooij1999}, which also uses superconducting loop structure, could work as more sensitive detector than a dc-SQUID, with the sensitivity below the standard quantum limit \cite{Ilichev2007}.
These attractive properties of superconducting loop structure pave the way towards realizing ultrasensitive EPR spectroscopy down to nanometer-sized area.

\begin{figure}[htbp]
\centering
\includegraphics[clip]{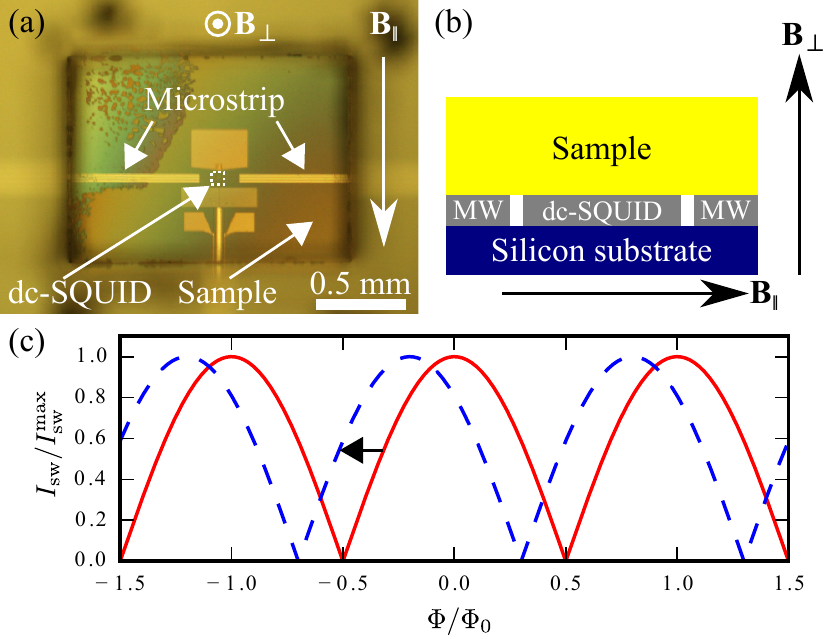}
\caption{
(color online.)
(a)
An optical microscope image of the device used in our experiment. The sample is directly bonded to the dc-SQUID fabricated on the silicon substrate.
(b)
A cross sectional view of the experimental setup. 
Two superconducting magnets are used for the experiment.
Smaller one is attached near the device to control the dc-SQUID ($\mathbf{B}_\perp$), and larger one is placed in the cryostat to generate the in-plane magnetic field ($\Bin$).
(c)
Switching current against magnetic flux penetrating the dc-SQUID.
Red solid (blue dash) line denotes the $I_{\mathrm{sw}}-\Phi$ curve of the dc-SQUID without (with) magnetization from an electron spin ensemble.
}
\label{fig:1}
\end{figure}

In Figs. \ref{fig:1}(a) and (b), we show our experimental setup.
We fabricate a dc-SQUID with a loop size of 26 $\times$ 7.25 $\mu$m on a silicon substrate.
Microstrips for excitation of the electron spin ensemble are also fabricated in the vicinity of the dc-SQUID. 
A sample that includes the electron spin ensemble is directly glued on the chip using vacuum grease.
It is worth mentioning that a similar technique has been used to couple a superconducting flux qubit with an ensemble of NV centers in diamond \cite{Zhu2011, Saito2013}.

In our method, the electron spin polarization is detected as follows \footnote{Characterization of superconducting material is performed using a similar method to that outlined in [\cite{Tsuchiya2014}].}:
Considering the compatibility with pulsed measurements and the readout scheme of the superconducting flux qubits, we measure the switching current of a dc-SQUID $I_{\mathrm{sw}}$ in time domain rather than the dc voltage.
This is determined by the magnetic flux penetrating the dc-SQUID $\Phi$:
\begin{equation}
	\frac{I_\mathrm{sw}}{I_\mathrm{sw}^\mathrm{max}} = \left| \cos \left( \pi \frac{\Phi}{\Phi_0}\right)\right|,
\end{equation}
where $I_\mathrm{sw}^\mathrm{max}$ is the maximum switching current;
$\Phi_0 = 2e/h$ is the magnetic flux quantum.
Further, $\Phi$ is the summation of magnetic flux generated by the superconducting magnet and the electron spin ensemble.
The polarization ratio of an electron spin ensemble depends on the ratio of the Zeeman and thermal energy.
We consider two extreme regimes as follows:
If thermal energy is dominant, e.g. under zero in-plane magnetic field, the electron spins are thermally fluctuated and there is no contribution to $\Phi$ [Fig. 1(c), red solid line].
On the other hand, if the Zeeman energy is dominant, e.g. under a finite in-plane field, the electron spin ensemble is partially polarized, which generates detectable magnetic flux penetrating the dc-SQUID.
If we apply a resonant microwave signal with this setup, we can detect the change of the electron spin polarization ratio from the change in switching current of the dc-SQUID [Fig. 1(c), blue dash line].

\begin{figure}[htbp]
\centering
\includegraphics[clip]{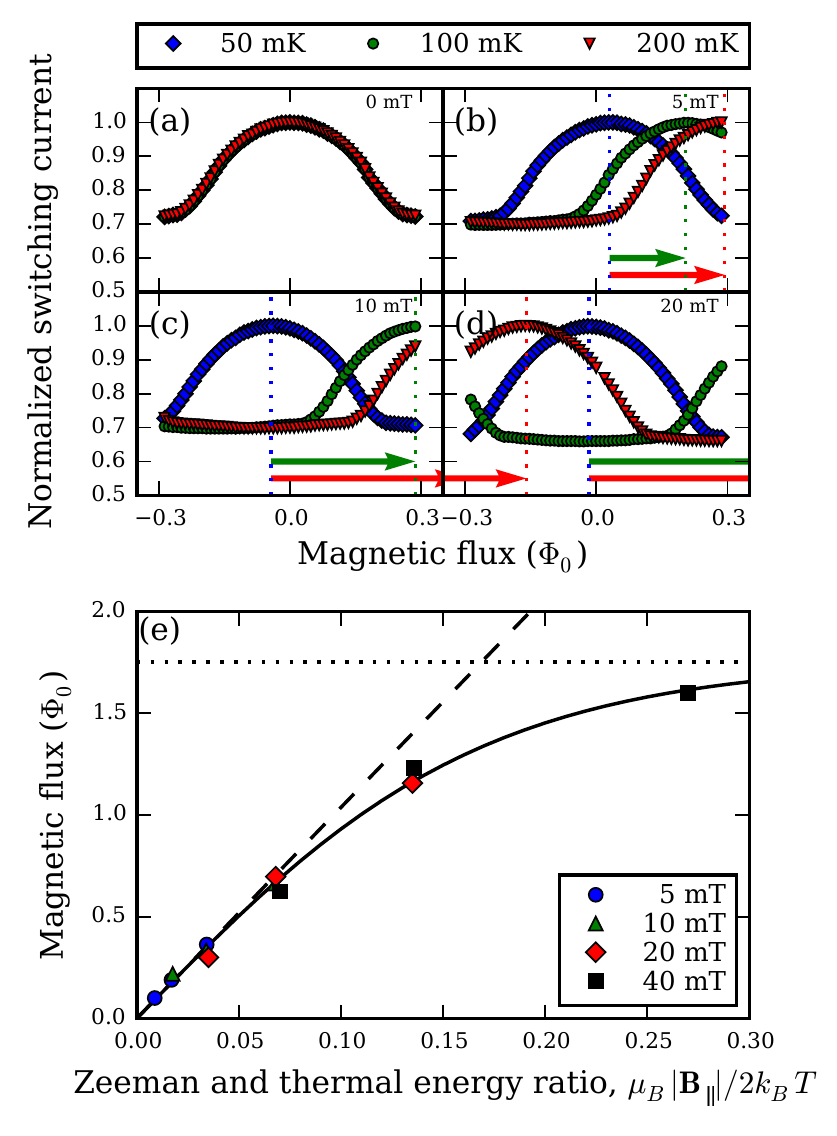}
\caption{
(color online.)
(a)--(d) External magnetic flux dependence of the switching current under various combinations of in-plane magnetic fields and temperatures.
(a) 0 mT,
(b) 5 mT,
(c) 10 mT,
and (d) 20 mT. 
The arrows appearing bottom of (b)--(d) represent the change of the magnetization from 50 mK.
It is worth mentioning that in (d), the modulation curve shift induced by the spin polarization approaches to one $\Phi_0$, thus the red curve seemingly has negative shift.
(e) Magnetization of the sample as a function of Zeeman and thermal energy ratio. Temperature of the sample stage is [50, 100, 200] mK.
Solid line is a fitting curve using Eq. (\ref{eq:2}).
Saturation flux (dotted line), an averaged Land\'{e} factor, and a flux offset are used as fitting parameters.
Dash line is a linear fitting curve using the slope at $|\Bin|=0$.
}
\label{fig:2}
\end{figure}

Let us now describe our experiment in a little more detail.
We perform a measurement of electron spin polarization without {the} microwave excitation.
For a proof-of-principle experiment, we use a single crystal YSO with 200 ppm of erbium dope (Scientific Materials, Inc.) as a spin ensemble.
It is crucial for the detection via the dc-SQUID that erbium impurities in a YSO crystal have an anisotropic Land\'e factor tensor.
Otherwise, an applied in-plane magnetic field would just change the polarization ratio of the electron spin with an in-plane direction, and this does not affect the amount of the magnetic flux penetrating the loop of the dc-SQUID.
Considering the Land\'e factor tensor, we choose {a} suitable alignment between the dc-SQUID, Er:YSO, and the in-plane magnetic field to maximize the vertical spin polarized component generated by the in-plane magnetic field \cite{Guillot-Noel2006, Sun2008}.
Measurements are performed in a ${}^3$He-${}^4$He dilution refrigerator whose base temperature is below 20 mK.
We use two superconducting magnets for the experiment [Fig. \ref{fig:1}(a)]: One is for controlling the dc-SQUID, and the other is for polarizing the electron spin ensemble.
To characterize the dc-SQUID, we use the switching method with 1000 times repetition.

In Figs. \ref{fig:2}(a)--(d), we plot the switching current against perpendicular magnetic flux generated by the superconductor magnet with different combinations of in-plane magnetic fields and temperatures
\footnote{Here, the switching current does not go down to zero possibly due to the asymmetric properties of the Josephson junctions.}.
Under zero in-plane magnetic field [Fig. \ref{fig:2}(a)], the electron spin ensemble is completely thermalized, thus the response of the dc-SQUID shows no temperature dependence.
On the other hand, under the finite magnetic field [Fig. \ref{fig:2}(b)--(d)], the electron spin ensemble is partially polarized and the amplitude of the curve shift depends on the temperature of the sample $T$.

{In Fig. \ref{fig:2}(e), we plot magnetic flux generated by electron spin polarization against a ratio of the Zeeman and thermal energy}
\footnote{We correct offsets between different $\Bin$ by assuming data that have {the} same Zeeman and thermal energy ratio generate same magnetic flux.}.
We observe a clear hyperbolic tangent dependence of the magnetic flux against the ratio, which indicates the polarization of the electron spin ensemble.
It is worth mentioning that especially in the high Zeeman energy regime ($\mu_B|\Bin|/2k_B T \gtrsim 0.13$), electron spin polarization saturates.
In general, the electron spin polarization ratio of two level systems has a hyperbolic tangent dependence as follows:
\begin{equation}
\tanh\left(\frac{\mu_B|\mathsf{g}\cdot \Bin|}{2k_B T}\right),
\label{eq:2}
\end{equation}
where $\mu_B$ is the Bohr magneton; $\mathsf{g}$ is the Land\'{e} factor tensor of the sample; $\Bin$ is the in-plane magnetic field; $k_B$ is Boltzmann constant. 
We use the saturation value of magnetic flux and the effective Land\'{e} factor $\tilde{g}$ as fitting parameters of Eq. (\ref{eq:2}), and these are estimated to be $1.75 \pm 0.21$ $\Phi_0 $ and $5.9 \pm 1.4$, respectively
\footnote{In this Letter, we use 1\% confidential intervals of the fitting constants as the error bars.}.
Here, we use an effective Land\'{e} factor $\tilde{g}$ instead of the tensor $\mathsf{g}$ for the following reason:
YSO crystal has two different crystallographic sites for erbium impurities.
The Land\'{e} factors for these two sites are different and they depend on the angle between a magnetic field and electron spin polarization.
Considering $C_2$ symmetry of the crystal, in total, there are four different Land\'{e} factor tensors for specific magnetic field configuration.
In our experiment, since we only measure the total magnetization of the electron spin ensemble, we can only get the averaged information of the sample, $\tilde{g}$.

We theoretically estimate the effective Land\'{e} factor $\tilde{g}$ for our experiment as follows:
Thermal population is calculated using the Boltzmann distribution $\exp(-E_k/k_BT)$.
Here, $E_k$ is an energy eigenvalue calculated from the spin Hamiltonian, 
\begin{equation}
H=\mu_B\Bin\cdot\mathsf{g}\cdot\mathbf{S}+\mathbf{I}\cdot \textsf{A}\cdot \mathbf{S}+\mathbf{I}\cdot\mathsf{Q}\cdot\mathbf{I},
\end{equation}
where $\mathbf{S}$ $(\mathbf{I})$ is a electron (nuclear) spin vector operator; $\textsf{A}$ is a hyperfine tensor; and $\mathsf{Q}$ is a nuclear quadruple tensor.
We can average the Land\'{e} factor due to the thermal population for different crystallographic sites or the zero field splitting induced by hyperfine interaction and so we can estimate $\tilde{g}$ {as} 6.2.
This value is consistent with the experimental result considering the confidential interval and angular misalignment of the magnetic field.

In general, the perpendicular component of the magnetization is needed to detect the spin polarization using our method, because the dc-SQUID detects the magnetic flux penetrating the loop.
In the case of Er:YSO based ensemble, the anisotropy of the Land\'{e} factor tensor allows us to use the in-plane magnetic field for generating the perpendicular magnetization.
To implement our spin detection scheme with spins having an isotropic Land\'{e} factor, we can put the spin-ensemble substrate that partially covers the dc-SQUID loop. In this case, by applying a in-plane magnetic field, the spin polarization induces magnetic flux penetrating the dc-SQUID loop.
Thus, the dc-SQUID based spin polarization detection scheme is widely applicable to any electron spin ensemble.

Let us now evaluate the sensitivity of our spin polarization detection scheme.
The minimum distinguishable magnetic flux is estimated to be $3.5 \times 10^{-3}\Phi_0$ by considering 1\% confidential intervals of the fitting parameter of the sinusoidal curves in Figs. \ref{fig:2}(a)--(d).
For these measurements, we use a spin ensemble whose concentration is $3.7\times 10^{18}$ $\mathrm{cm}^{-3}$
and the saturation flux is estimated to be 1.75 $\Phi_0$.
From these values, the minimum distinguishable spin concentration is estimated to be $7.4\pm1.0\times 10^{15}$ $\mathrm{cm}^{-3}$.

We can also estimate the sensing volume.
The sensing area is limited in the dc-SQUID loop.
In our case, it is 101 $\mu$m$^2$.
It is worth mentioning that compared to conventional EPR spectrometers, our method is more sensitive to the spin ensemble near the interface between the sample and the dc-SQUID. 
This is because the magnetic coupling strength between them is determined by the distance.
Assuming a few micrometer of effective thickness
\footnote{In the case of a flux qubit, theoretical calculations show that coupling strength between the spin ensemble and the flux qubit decays rapidly as a function of the distance between them, especially if the distance is larger than 1 $\mu$m. See Ref. \cite{Marcos2010}.}, the effective sensing volume of this experiment is estimated to be $\sim$$10^{-10}$ $\mathrm{cm}^{3}$ ($\sim$0.1 pl), which is 100 times smaller than the ESR spectroscopy using a lamped element on-chip resonator \cite{Bienfait2015}. 
Considering this sensing volume, the minimum distinguishable number of electron spins detected by the dc-SQUID is around $10^{6}$.

\begin{figure}[htbp]
\centering
\includegraphics[clip]{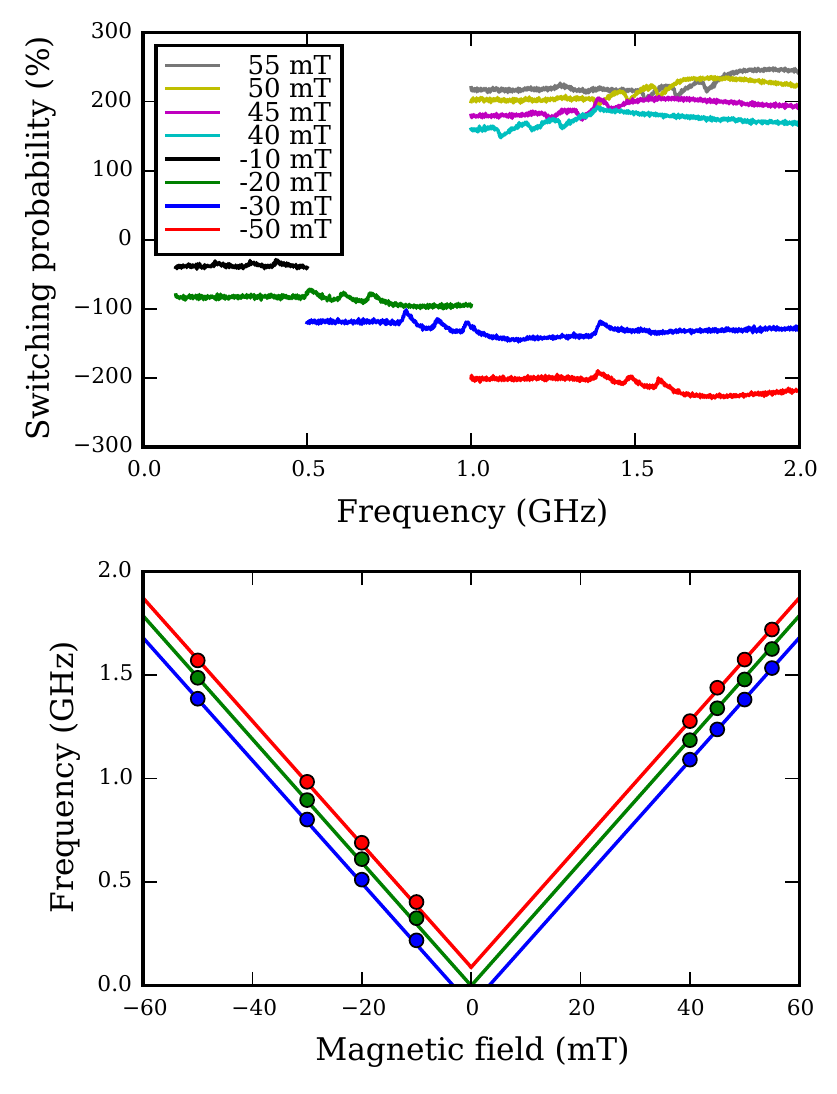}
\caption{(color online.) 
Electron paramagnetic resonance spectroscopy of diamond crystal. 
(a) The relative switching probability as a function of the microwave excitation frequency. 
Lines are shifted by $4 \times \Bin/\mathrm{mT}$ (\%) for clarity. (b) Magnetic field dependence of the resonance frequency.}
\label{fig:3}
\end{figure}

Figs. \ref{fig:3} shows the results of our EPR spectroscopy for a type Ib (100) diamond.
In this experiment, the in-plane magnetic field is applied along the [110] direction.
The perpendicular magnetic field is fixed at some bias point, and {the} change in the switching probability is recorded as a signal.
The experiment is performed at the base temperature of the refrigerator to maximize the spin polarization ratio.
To excite the electron spin ensemble, {a} continuous microwave signal is applied to the sample through one of the microstrips.

Fig. \ref{fig:3}(a) shows the results of EPR spectroscopy in frequency sweep experiments.
Clear resonance peaks are observed.
In addition to the center peaks, two satellite peaks shifted by $\sim$100 MHz are also observed.
These peaks have an asymmetry in shape arising from thermal effects.
The electron spins are energetically excited by microwave radiation, whose energy is transferred to the lattice system through energy relaxation process.
At low temperature, since thermal conductivity is significantly reduced, the energy transfer process from the lattice to the cold stage become a bottleneck of the thermal transport, which increases the effective sample temperature.
To examine this effect, we also performed EPR spectroscopy using the Er:YSO crystal.
In this case, the asymmetry of the resonance peaks is much larger than the case of diamond, despite using short pulsed microwave excitation.
The origin of the difference is thought to be the effect of different thermal conductivity:
Diamond has about a 100 times better thermal conductivity than YSO.
This large difference results in the asymmetry difference.

In Fig. \ref{fig:3}(b), we plot the resonant frequency observed in the ESR experiment against the in-plane magnetic field.
From a linear fitting, the Land\'{e} factor and peak spacing are estimated to be 2.12 $\pm$ 0.02 and 93 $\pm$ 14 MHz, respectively.
Considering the calibration error of the superconducting magnet, these values indicate that the peaks observed in Fig. \ref{fig:3} (b) originate from P1 centers in diamond\cite{Smith1959}.
Due to the hyperfine interaction of $^{14}$N ($I=1$), we observe three peaks: 
main peaks (green) arise without the contribution from the nuclear spin ($I_z=0$), and two satellite peaks (blue and red) have hyperfine splitting ($I_z=\pm1$).
The splitting width of the satellite peaks depends on the angle between the nitrogen-carbon (N-C) bond and electron spin polarization.
In this magnetic field configuration, two N-C bonds are perpendicular to the in-plane field, and the others make the angle of 35.3 degree.
Although the satellite peaks could have further splitting corresponding to these two angles, measured bandwidth $\sim$40 MHz is too broad to observe this splitting.

It is worth mentioning that the effective sensing volume of the EPR spectroscopy is smaller than that of the spin polarization detection.
This is because the microwave excitation power is strong enough only in the vicinity of the microstrip.
Thus, although it is difficult to define an exact sensing volume, we can expect this is estimated to be much less than $\sim$$10^{-10}$ $\mathrm{cm}^{3}$ ($\sim$0.1 pl)
\footnote{We check this effect by using two microstrips whose distance from the dc-SQUID is different. In the case of 4 $\mu$m one, we successfully confirm EPR spectrum, although in the case of 9 $\mu$m, we cannot confirm EPR.}.

In conclusion, we have successfully performed EPR spectroscopy by using a dc-SQUID magnetometer as a detector of electron spin polarization.
The minimum distinguishable number of polarized spins and sensing volume are estimated to be $\sim$$10^6$ and $\sim$$10^{-10}$ $\mathrm{cm}^{3}$ ($\sim$0.1 pl), respectively.
This sensing volume is 100 times smaller than the ESR spectrometer using a lamped element on-chip resonator.
We can in principle control its spatial resolution and sensitivity by changing loop size of a dc-SQUID, which is applicable for nanoscale characterization of materials.

This work was supported by Commissioned Research of NICT and in part by MEXT Grant-in-Aid for Scientific Research on
Innovative Areas ``Science of hybrid quantum systems'' (Grant No. 15648489 and 15H05869).


\end{document}